\documentclass[a4paper,10pt,twoside]{cpc-hepnp}

\usepackage{multicol}
\usepackage{graphicx}
\usepackage{booktabs}
\usepackage{amssymb,bm,mathrsfs,bbm,amscd}
\usepackage[tbtags]{amsmath}
\usepackage{lastpage}

\begin{document}
%\begin{CJK*}{GBK}{song}

\fancyhead[co]{\footnotesize Huang Ming-Yang et al.: Detection of supernova neutrinos at spallation neutron sources}

%\footnotetext[0]{Received 14 March 2009}

\title{Detection of supernova neutrinos at spallation neutron sources\thanks{Supported by National Natural
Science Foundation of China (11205185, 11175020, 11275025, and 11575023)}}

\author{%
      Huang Ming-Yang$^{1,2;1)}$\email{huangmy@ihep.ac.cn}%
 \quad GUO Xin-Heng$^{3;2)}$\email{xhguo@bnu.edu.cn}%
 \quad YOUNG Bing-Lin$^{4,5;3)}$\email{young@iastate.edu}%
}
\maketitle

\address{%
$^1$ Institute of High Energy Physics (IHEP), Chinese Academy of Sciences (CAS),
Beijing 100049, China\\
$^2$ Dongguan Institute of Neutron Science (DINS), Dongguan 523808, China\\
$^3$ College of Nuclear Science and Technology, Beijing Normal University,
Beijing 100875, China\\
$^4$ Department of Physics and Astronomy, Iowa State University, Ames, Iowa
5001, USA \\
$^5$ Institute of Theoretical Physics, Chinese Academy of Sciences, Beijing,
China\\
}

\begin{abstract}

After considering the supernova shock effects, the Mikheyev-Smirnov-Wolfenstein effects,
the neutrino collective effects, and the Earth matter
effects, the detection of supernova neutrinos at China Spallation Neutron Sources
is studied and the event numbers of different flavor supernova neutrinos
observed through various reaction channels are calculated with the neutrino energy
spectra described by the Fermi-Dirac distribution and ``beta fit" distribution
respectively. Furthermore, the numerical calculation method of supernova neutrino detection on the Earth
is applied to some other spallation neutron sources,
and the total event numbers of supernova neutrinos observed through different reactions
channels are given.
\end{abstract}



%\footnotetext[0]{\hspace*{-3mm}\raisebox{0.3ex}{$\scriptstyle\copyright$}2013
%Chinese Physical Society and the Institute of High Energy Physics
%of the Chinese Academy of Sciences and the Institute
%of Modern Physics of the Chinese Academy of Sciences and IOP Publishing Ltd}%

\begin{multicols}{2}

\section{Introduction}

Supernovas (SNs) are extremely powerful explosions in
the universe which terminate the life of some stars.
They make the catastrophic end of stars more massive than
eight solar masses ($>8M_{\odot}$), leaving behind compact remnants such
as neutron stars or black holes. The SN explosion is one of the most spectacular cosmic events and
a source of new physical ideas \cite{Mirizzi1}. A broad area of
fundamental physics can be studied by the observation of SN \cite{Kotake}.
Detection of SN neutrinos on the Earth \cite{Guo1}, such as SN1987A \cite{SN19871, SN19872},
has been a subject of intense investigation in astroparticle physics.
Some information about the SN explosion mechanism
and neutrino mixing parameters can be obtained by detecting SN neutrinos
on the Earth \cite{Kotake, Scholberg}.

China Spallation Neutron Source (CSNS) \cite{CSNS2} is a high power accelerator
based facility. It consists of an 80 MeV proton linac,
a 1.6 GeV Rapid Cycling Synchrotron (RCS), a solid tungsten target station, and instruments
for spallation neutron applications \cite{Wei3}. The accelerator operates at 25 HZ repetition rate with
an initial design beam power of 100 kW and is upgradeable to 500 kW \cite{Wang1,Huang2}.
As the exclusive spallation neutron source in developing countries, CSNS will be among the top four of such facilities in the world until completion.

\begin{center}
\includegraphics[width=9cm]{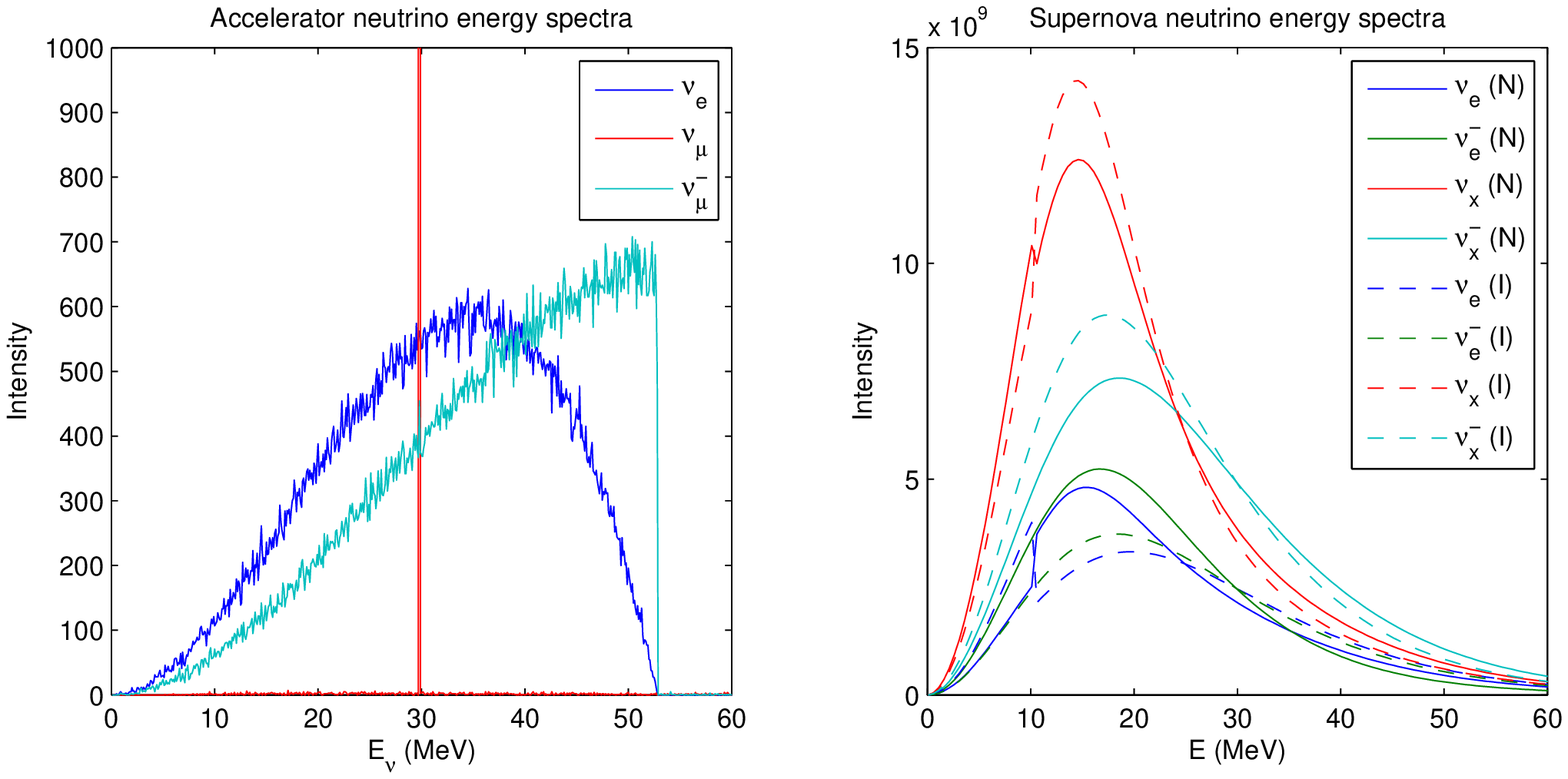}
\figcaption{\label{fig1} Neutrino energy spectra. (a) accelerator neutrinos at CSNS;
(b) SN neutrinos on the Earth. $N (I)$ corresponds to the normal (inverted) mass hierarchy, and $x=\mu, \tau$.
}
\end{center}

By using the code FLUKA, the processes of accelerator neutrinos production during the proton
beam hitting on the tungsten target at CSNS were simulated, and the energy
spectra of accelerator neutrinos were gained \cite{Huang5}, as shown in Fig. 1(a). While
considering the SN shock effects \cite{Wilson1, Schirato1, Hudepohl1, Sarikas1},
the Mikheyev-Smirnov-Wolfenstein (MSW) effects \cite{MSW1, MSW2, Kuo1, Blennow}, the
neutrino collective effects \cite{Dasgupta1, Duan1, Duan2}, and the Earth matter
effects \cite{Dighe2, Ioannisian1, Guo2}, the detection of SN neutrinos on the Earth
was studied \cite{Huang1}. Then, the energy spectra of SN neutrinos can be calculated, as shown in Fig. 1(b).
To compare the energy spectra of accelerator neutrinos with that of SN neutrinos, it is clear that the
energy spectrum ranges of SN neutrinos are very close to that
of accelerator neutrinos. Therefore, by using the accelerator neutrino detector at spallation neutron sources,
different flavor neutrinos from a SN explosion can also be detected,
and then can be served as a SN Early Warning System \cite{Antonioli}.

\section{SN neutrino detection at CSNS}

In the core collapse of SN, a vast amount of neutrinos are produced
in two bursts \cite{book1, book2}. When SN neutrinos of each flavor are produced,
they are approximately the effective mass eigenstates due to the extremely
high matter density environment. While neutrinos propagate outward to the surface of the SN, they
could be subjected to the SN shock effects,
the MSW effects, and the neutrino collective effects.
Then, after travelling the cosmic distance to reach the Earth, they
go through a part of the Earth and are subjected to the Earth matter
effects. Fig. 2 shows the path of SN neutrinos reaching the detector on the Earth.

\begin{center}
\includegraphics[width=6cm]{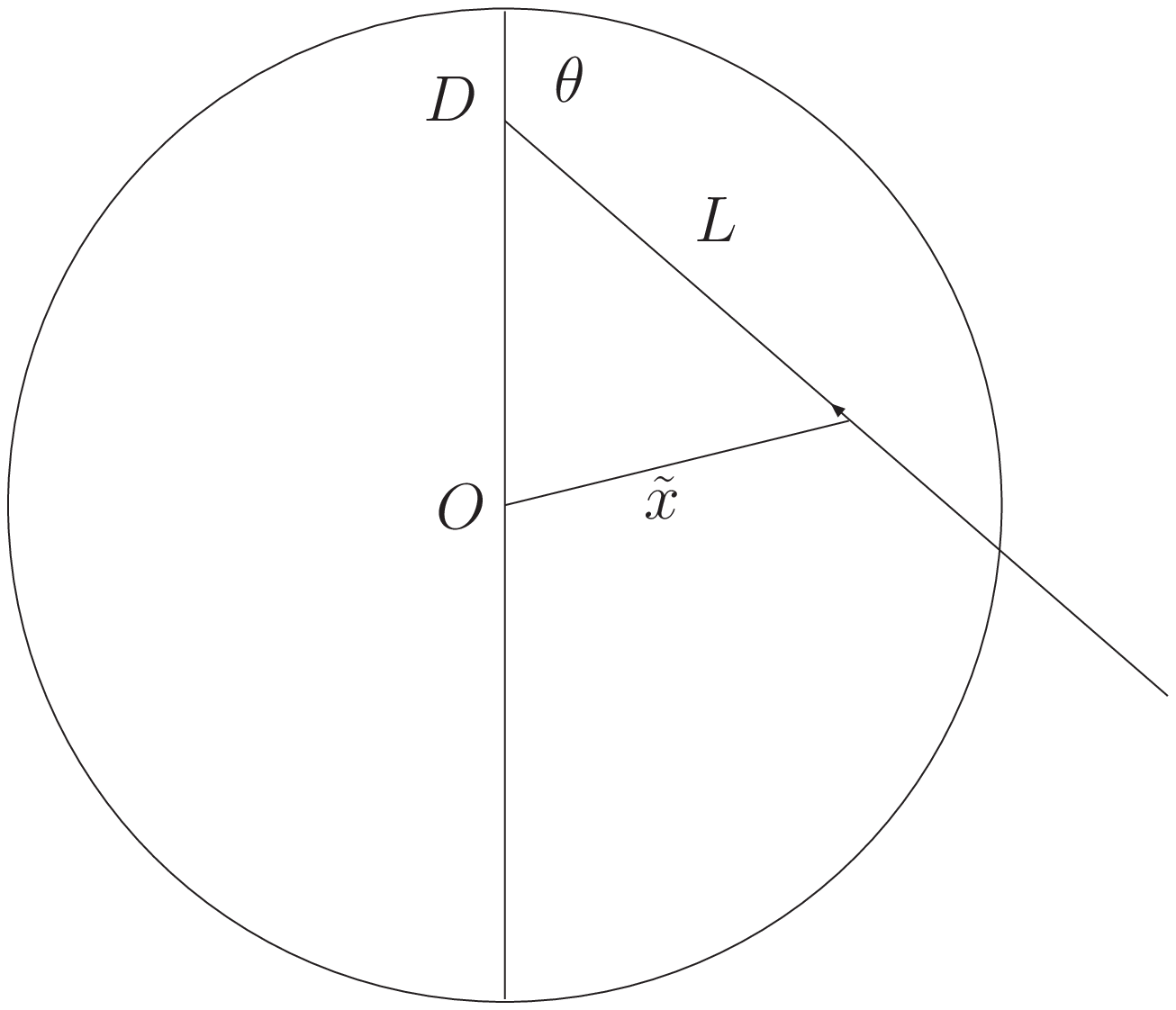}
\figcaption{\label{fig2} Illustration of the path of SN neutrinos
reaching the detector in the Earth. $D$ is the location of the
detector, $\theta$ is the incident angle of neutrinos,
$O$ is the center of the Earth, $L$ is the distance neutrinos
travel through the Earth, and $\tilde{x}$ is the distance of
neutrinos to the center of the Earth.}
\end{center}

When all effects, including the SN shock effects, the MSW effects, the neutrino collective effects,
and the Earth matter effects, are taken into account, the SN neutrino fluxes at the
detector can be written as \cite{Huang1}
\begin{eqnarray}
F_{\nu_e}^D&=&pF_{\nu_e}^{(0)}+(1-p)F_{\nu_x}^{(0)},
\nonumber \\
F_{\bar{\nu}_e}^D&=&\bar{p}F_{\bar{\nu}_e}^{(0)}+
  (1-\bar{p})F_{\bar{\nu}_x}^{(0)},
\nonumber \\
2F_{\nu_x}^D&=&(1-p)F_{\nu_e}^{(0)}+(1+p)F_{\nu_x}^{(0)},
\nonumber \\
2F_{\bar{\nu}_x}^D &=&(1-\bar{p})F_{\bar{\nu}_e}^{(0)}+
 (1+\bar{p})F_{\bar{\nu}_x}^{(0)},
\label{FD}
\end{eqnarray}
where $x=\mu, \tau$, and the survival probabilities $p$ and $\bar{p}$ are given by
\begin{eqnarray}
p&=&P_{2e}[P_HP_{\nu\nu}+(1-P_H)(1-P_{\nu\nu})], \nonumber\\
\bar{p}&=&(1-\bar{P}_{2e})\bar{P}_{\nu\nu}, \label{pn}
\end{eqnarray}
for the normal mass hierarchy and
\begin{eqnarray}
p&=&P_{2e}P_{\nu\nu}, \nonumber\\
\bar{p}&=&(1-\bar{P}_{2e})[\bar{P}_H\bar{P}_{\nu\nu}+
  (1-\bar{P}_H)(1-\bar{P}_{\nu\nu})],
\label{pi}
\end{eqnarray}
for the inverted mass hierarchy. In Eqs. (\ref{pn}) and (\ref{pi}), $P_{2e}$($\bar{P}_{2e}$)
is the probability that a (anti)neutrino mass eigenstate
$\nu_2$($\bar{\nu}_2$) enters the surface of the Earth and arrives at the detector
as an electron (anti)neutrino $\nu_e$($\bar{\nu}_e$), $P_{\nu\nu}$($\bar{P}_{\nu\nu}$)
is the probability that the (anti)neutrino $\nu$($\bar{\nu}$)
remains as $\nu$($\bar{\nu}$) after the collective effects,
and $P_H$($\bar{P}_H$) is the crossing probability for (anti)neutrinos to jump from
one eigenstate to another at the high resonance layer \cite{Huang1}.

We assume a ``standard" SN explosion at a distance $D=10$ kpc from the
Earth, releasing a total energy $E_B=3\times10^{53}$ erg (similar to
SN1987A \cite{SN19871, SN19872}). For the SN neutrino of flavor $\alpha$ ($\alpha=e, \mu, \tau$),
the luminosity flux is distributed in time as
\begin{equation}
 L_{\alpha}(t)=\frac{E_B}{18}\exp(-t/3). \label{Lum}
\end{equation}
In general simulations, the time-integrated neutrino
energy spectra can be described by the Fermi-Dirca distribution (the ``Livermore" model) \cite{Totani1}
or ``beta fit" distribution (the ``Garching" model) \cite{Keil5}:

(1) Fermi-Dirca distribution
\begin{equation}
 F_{\alpha}^{(0)}(E)=\frac{L_{\alpha}(t)}{F_{\alpha
 3}T_{\alpha}^{4}}\frac{E^2}{\exp{(E/T_{\alpha}-\eta_\alpha)}+1},
 \label{Foa}
\end{equation}
where $E$ is the neutrino energy, $T_{\alpha}$ is the temperature of
the neutrino $\alpha$, $\eta_\alpha$ is the dimensionless pinching parameter of the
spectrum, and
$F_{\alpha j}$ is defined by
\begin{equation}
 F_{\alpha j}=\int_{0}^{\infty}\frac{x^j}{\exp{(x-\eta_\alpha)}+1}{\rm
 d}x, \nonumber
\end{equation}
where $j$ is an integer. The spectra obtained from numerical
simulations can be well fitted by \cite{Janka1, Janka2}
\begin{eqnarray}
 T_{\nu_e}=3-4MeV,& \eta_{\nu_e}\approx3-5, \nonumber\\
 T_{\bar{\nu}_e}=5-6MeV,& \eta_{\bar{\nu}_e}\approx2.0-2.5, \label{T} \\
 T_{\nu_x}=T_{\bar{\nu}_x}=7-9MeV,& \eta_{\nu_x}=\eta_{\bar{\nu}_x}\approx0-2. \nonumber
 \end{eqnarray}

(2) ``Beta fit" distribution
\begin{eqnarray}
 F_{\alpha}^{(0)}(E)=\frac{L_{\alpha}(t)}{\langle E_{\alpha}\rangle^2}\frac{\beta_{\alpha}^{\beta_{\alpha}}}{\Gamma(\beta_{\alpha})}
 \Bigg(\frac{E_{\alpha}}{\langle E_{\alpha}\rangle}\Bigg)^{\beta_{\alpha}-1} \nonumber \\
 \times\exp\Bigg(-\beta_{\alpha}\frac{E_{\alpha}}{\langle E_{\alpha}\rangle}\Bigg),
 \label{Foa2}
\end{eqnarray}
where $\langle E_{\alpha}\rangle$ is the neutrino average energy and $\beta_{\alpha}$ is
the dimensionless pinching parameter. The spectra obtained from numerical
simulations can be well fitted by \cite{Kotake, Chakraborty5}
\begin{eqnarray}
  && \langle E_{\nu_e}\rangle=\langle E_{\bar{\nu}_e}\rangle = 12\sim15 MeV,
  \label{E} \\
  && \langle E_{\nu_x}\rangle=\langle E_{\bar{\nu}_x}\rangle = 15\sim18 MeV, \quad \beta_{\alpha} = 3.5\sim6. \nonumber
 \end{eqnarray}

Therefore, the event numbers $N(i)$ of SN neutrinos observed through various
reaction channels ``$i$" can be calculated by
\begin{equation}
 N(i)=N_T\int{{\rm d}E\cdot\sigma(i)\cdot\frac{1}{4\pi
 D^2}\cdot F_{\alpha}^D}, \label{Ntotal}
\end{equation}
where $N_T$ is the target number, $\sigma(i)$ is the cross section
of the given reaction channel, and $D$ is the distance between the SN and the Earth.

For CSNS \cite{Huang5}, a medium scale detector would be placed just below the ground
surface about 50-60 meters from the spallation target. A spherical
803 tons fiducial mass of mineral oil (CH$_2$, density 0.845 g/cm$^3$) has a fiducial radius
of 6.1 m, occupying a volume of 950 m$^3$.
Then the total event numbers of target protons, electrons, and $^{12}C$ are
\begin{eqnarray}
&&\ N_T^{(p)}=6.90\times10^{31},
\quad N_T^{(e)}=2.76\times10^{32}, \nonumber \\
&&\ N_T^{(C)}=3.45\times10^{31}. \nonumber
\end{eqnarray}

\end{multicols}

\begin{center}
\tabcaption{ \label{tab1}  Reaction channels used to detect SN neutrinos at CSNS,
where $E_{th}$ is the reaction threshold, the unit of $E$ is MeV,
and the cross sections for the $^{12}C$ reactions are the average cross sections.}
\footnotesize
\begin{tabular*}{170mm}{@{\extracolsep{\fill}}c|cccc}
\toprule
Reaction  & Equation   &$E_{th}$ (MeV) & Target numbers & Cross sections (cm$^2$)  \\
 \hline
 $\bar{\nu}_ep$   &   $\bar{\nu}_e+p\rightarrow e^++n$   &   1.8    &    $6.9\times10^{31}$  &  $9.5\times10^{-44}(E-1.29)^2 $ \\
 \hline
    &   $\nu_e+e^-\rightarrow \nu_e+e^-$  &   0    &    $2.76\times10^{32}$
    & $9.20\times10^{-45}E$ \\
 $\nu e^-$  &   $\bar{\nu}_e+e^-\rightarrow \bar{\nu}_e+e^-$  &   0    &    $2.76\times10^{32}$
 &  $3.83\times10^{-45}E$  \\
 &   $\nu_x+e^-\rightarrow \nu_x+e^-$  &    0  &   $2.76\times10^{32}$
 & $1.57\times10^{-45}E$   \\
 &   $\bar{\nu}_x+e^-\rightarrow \bar{\nu}_x+e^-$  &   0   &  $2.76\times10^{32}$
 & $1.29\times10^{-45}E$    \\
 \hline
 &   $\nu_e+^{12}C\rightarrow^{12}N+e^-$  &   17.34   &    $3.45\times10^{31}$  & $1.85\times10^{-43}$  \\
 $\nu^{12}C$ &   $\bar{\nu}_e+^{12}C\rightarrow^{12}B+e^+$  &   14.39   &    $3.45\times10^{31}$
 & $1.87\times10^{-42}$   \\
 &   $\nu_e+^{12}C\rightarrow^{12}C^*+\nu'_e$  &   15.11   &   $3.45\times10^{31}$
 & $1.33\times10^{-43}$  \\
 &   $\bar{\nu}_e+^{12}C\rightarrow^{12}C^*+\bar{\nu}'_e$  &   15.11   &   $3.45\times10^{31}$
 &  $6.88\times10^{-43}$\\
 &   $\nu_x+^{12}C\rightarrow^{12}C^*+\nu'_x$  &   15.11   &   $3.45\times10^{31}$
 & $3.73\times10^{-42}$  \\
 &   $\bar{\nu}_x+^{12}C\rightarrow^{12}C^*+\bar{\nu}'_x$  &   15.11   &   $3.45\times10^{31}$
 & $3.73\times10^{-42}$ \\
\bottomrule
\end{tabular*}%
\end{center}

\begin{multicols}{2}

It is clear that there are three reaction channels which can be
used to detect SN neutrinos: the inverse beta decay,
the neutrino-electron reactions, and the neutrino-carbon reactions.
Table 1 shows the reaction thresholds, target numbers, and effective cross sections
for the three reactions \cite{Cadonati1, Arafune1, Fukugita1, Kolbe5}.
It can be seen that, for the inverse beta decay, the neutrino events can be
identified by the detection of both the $e^+$ and the 2.2 MeV $\gamma$ from the
reaction $n+p\rightarrow d+\gamma$ with a mean capture time $\tau=250\mu s$ \cite{Athanassopulos1, Athanassopulos2};
for the neutrino-electron reactions, the neutrino events can be identified
by the signal of the recoil electrons which are strong peaked along the neutrino direction,
and this forward peaking is usually used for experiments to distinguish the electron elastic scattering
from the neutrino reactions on nuclei \cite{Allen2, Imlay1}; for the neutrino reactions on $^{12}C$, there are two charged-current and six neutral-current reactions:

Charged-current capture of $\nu_e$:
\begin{eqnarray}
&&\ \nu_e+^{12}C\rightarrow^{12}N+e^-, \quad E_{th}=17.34MeV,
\nonumber
\\
&&\ ^{12}N\rightarrow^{12}C+e^{+}+\nu_e, \quad \tau_{1/2}=11.00 ms.
\nonumber
\end{eqnarray}

Charged-current capture of $\bar{\nu}_e$:
\begin{eqnarray}
&&\ \bar{\nu}_e+^{12}C\rightarrow^{12}B+e^+, \quad E_{th}=14.39MeV,
\nonumber
\\
&&\ ^{12}B\rightarrow^{12}C+e^{-}+\bar{\nu}_e, \quad \tau_{1/2}=20.20 ms.
\nonumber
\end{eqnarray}

Neutral-current inelastic scattering of $\nu_{\alpha}$ or $\bar{\nu}_{\alpha}$ ($\alpha=e, \mu, \tau$):
\begin{eqnarray}
&&\ \nu_{\alpha}(\bar{\nu}_{\alpha})+^{12}C\rightarrow^{12}C^{\ast}+\nu_{\alpha}^{'}(\bar{\nu}_{\alpha}^{'}),
\quad E_{th}=15.11MeV,
\nonumber \\
 &&\ ^{12}C^{\ast}\rightarrow^{12}C+\gamma.
\nonumber
 \end{eqnarray}

The charged-current events have the delayed coincidence of a $\beta$ decay following the interaction.
The neutral-current events have a monoenergetic $\gamma$ ray at $15.11$ MeV. Therefore,
the charged-current and neutral-current reactions on carbon can be tagged and observed by the neutrino detector
\cite{Cadonati1, Auerbach1}.

In Table 1, for the neutrino-carbon reactions, the effective cross sections of
the charged-current interaction are given for
SN neutrinos without oscillations. When neutrino oscillations are taken into account, the oscillations of higher
energy $\nu_x$ into $\nu_e$ result in an increasing event rate
since the expected $\nu_e$ energies are just at or below the
reaction threshold. This leads to an increase by a
factor of 35 for the efficiency cross section
$\langle\sigma(^{12}C(\nu_e,e^-)^{12}N)\rangle$.
Similarly, the efficiency cross section
$\langle\sigma(^{12}C(\Bar{\nu}_e,e^+)^{12}B)\rangle$ increases
by a factor of 5.

Since the energy spectrum ranges of accelerator neutrinos and reactor neutrinos are
close to that of SN neutrinos, the background due to the accelerator neutrinos from CSNS
and the reactor neutrinos from Daya Bay reactor need to be estimated while observing SN neutrinos.
After careful calculation and analysis \cite{Huang5}, with the neutrino detector at CSNS, the event number
of accelerator neutrinos which can be observed is about $10^{-3}\thicksim10^{-4}$ per second.
The neutrino luminosity of Daya Bay reactor is very large \cite{DYB}, however, due to the long distance of
Daya Bay reactor from CSNS (about 70 km), the event number of reactor neutrinos
observed at CSNS is about $10^{-3}\thicksim10^{-4}$ per second by detailed calculation. It is
known that the SN explosion lasts for only about 20 second. Therefore, the event numbers
of accelerator neutrinos and reactor neutrinos are very few and can be ignored during the detection of
SN neutrinos at CSNS.

SN relic neutrinos, also known as diffuse SN neutrino background,
is of intense interest in neutrino astronomy and neutrino physics \cite{Nakazato2}.
With the neutrino detector at CSNS, the event number of SN relic neutrinos which can be observed \cite{Zhang2} is about $10^{-7}\thicksim10^{-8}$ per second. Due to the very short time of SN explosion,
the background of SN relic neutrinos can also be neglected.

In general, there is no serious background because of
the characteristics of the SN neutrino events which are concentrated in a short 20 second interval
with energies no more than 30 MeV. This has been confirmed in the Kamiokande \cite{SN19871} and
IMB \cite{SN19872} neutrino events of SN1987A.

In order to calculate the event numbers of SN neutrinos more accurately, the energy resolution and event
selection need to be studied, and then the detector efficiency can be obtained. For the proposed CSNS detector,
we choose to use the detector efficiency $\varepsilon_{pc}$ for the inverse beta decay, $\varepsilon_{ec}$ for the neutrino-electron reactions, $\varepsilon_{cc}$ for the neutrino-carbon reactions, respectively, during the
calculation of the SN neutrino event numbers \cite{VanDalen1}.

To make use of the latest experiment results of neutrino oscillations \cite{Olive1, Gapozzi1},
the neutrino mixing parameters are given:
\begin{eqnarray}
\ \Delta m^2_{21}=7.5\times10^{-5} eV^2,
\quad \mid\Delta m^2_{32}\mid=2.4\times10^{-3} eV^2, \nonumber \\
\ \sin^2\theta_{12}=0.308, \quad \sin^2\theta_{23}=0.446, \quad \sin^2\theta_{13}=0.0237. \nonumber
\end{eqnarray}

\end{multicols}

\begin{center}
\tabcaption{ \label{tab2} Summary of the event number ranges of different flavor
SN neutrinos detected in various reaction channels at CSNS. ``Range (FD)" (``Range (BF)")
stands for the event number ranges of SN neutrinos calculated by
using the Fermi-Dirac (``beta fit") distribution, ``[x, y]$\varepsilon$" stands for the event number
range satisfied $x\times\varepsilon\leqslant N \leqslant y\times\varepsilon$ where $\varepsilon$ is the
detector efficiency. }
\footnotesize
\begin{tabular*}{170mm}{@{\extracolsep{\fill}}c|cccccc}
\toprule Hierarchy  & Reaction   & Flavor    & Range (FD)  & Range (BF)  \\
  \hline
           &   $\bar{\nu}_ep$        &  $\bar{\nu}_e$         &    [370.81, 512.94]$\varepsilon_{pc}$ &
           [221.15, 316.95]$\varepsilon_{pc}$ \\
           &   $\nu e^-$             &  $\nu_e$   &                [6.36, 6.72]$\varepsilon_{ec}$       &
           [6.62, 6.73]$\varepsilon_{ec}$  \\
           &                         &  $\bar{\nu}_e$         &    [2.76, 2.83]$\varepsilon_{ec}$       &
           [2.81, 2.83]$\varepsilon_{ec}$    \\
           &                         &  $\nu_x$         &          [2.34, 2.40]$\varepsilon_{ec}$        &
           [2.33, 2.35]$\varepsilon_{ec}$        \\
 Normal    &                         &  $\bar{\nu}_x$         &    [1.91, 1.93]$\varepsilon_{ec}$        &
           [1.91, 1.92]$\varepsilon_{ec}$     \\
           &    $\nu^{12}C$          &  $\nu_e$         &          [21.69, 34.53]$\varepsilon_{cc}$      &
           [41.52, 52.35]$\varepsilon_{cc}$     \\
           &                         &  $\bar{\nu}_e$         &    [16.61, 25.15]$\varepsilon_{cc}$      &
           [29.74, 37.29]$\varepsilon_{cc}$  \\
           &                         &  $\nu_x$         &          [13.48, 20.01]$\varepsilon_{cc}$      &
           [23.68, 28.08]$\varepsilon_{cc}$    \\
           &                         &  $\bar{\nu}_x$         &    [18.99, 28.13]$\varepsilon_{cc}$      &
           [34.00, 41.11]$\varepsilon_{cc}$    \\
 \hline
           &   $\bar{\nu}_ep$        &  $\bar{\nu}_e$         &   [446.31, 640.79]$\varepsilon_{pc}$     &
           [255.22, 354.61]$\varepsilon_{pc}$  \\
           &   $\nu e^-$             &  $\nu_e$   &               [6.86, 7.21]$\varepsilon_{ec}$         &
           [6.92, 7.08]$\varepsilon_{ec}$ \\
           &                         &  $\bar{\nu}_e$         &   [2.85, 2.87]$\varepsilon_{ec}$         &
           [2.84, 2.85]$\varepsilon_{ec}$   \\
           &                         &  $\nu_x$         &         [2.25, 2.31]$\varepsilon_{ec}$         &
           [2.28, 2.30]$\varepsilon_{ec}$    \\
 Inverted  &                         &  $\bar{\nu}_x$         &   [1.89, 1.90]$\varepsilon_{ec}$         &
           [1.90, 1.91]$\varepsilon_{ec}$      \\
           &    $\nu^{12}C$          &  $\nu_e$         &         [28.77, 41.21]$\varepsilon_{cc}$       &
           [49.20, 58.91]$\varepsilon_{cc}$   \\
           &                         &  $\bar{\nu}_e$         &   [33.10, 48.19]$\varepsilon_{cc}$       &
           [58.52, 70.33]$\varepsilon_{cc}$   \\
           &                         &  $\nu_x$         &         [11.35, 17.21]$\varepsilon_{cc}$       &
           [21.13, 25.45]$\varepsilon_{cc}$   \\
           &                         &  $\bar{\nu}_x$         &   [14.55, 21.15]$\varepsilon_{cc}$       &
           [25.85, 31.32]$\varepsilon_{cc}$  \\
\bottomrule
\end{tabular*}%
\end{center}

\begin{multicols}{2}

\vspace{0.3cm}

By using Eqs. (\ref{FD})-(\ref{Ntotal}), the event numbers of SN neutrinos detected
on the Earth can be calculated. The numerical results calculated with the neutrino energy
spectra described by the Fermi-Dirac
distribution and ``beta fit" distribution are both shown in Table 2.
In the table, the event number ranges of different flavor neutrinos observed through
various reaction channels: the inverse beta decay, the neutrino-electron reactions,
and the neutrino-carbon reactions, are given. It can be found that:

(i) If $\varepsilon_{pc}\simeq\varepsilon_{ec}\simeq\varepsilon_{cc}$,
the total event number of different flavor neutrinos observed through the channel
of the neutrino-electron reactions is much smaller than that of
the inverse beta decay and neutrino-carbon reactions;

(ii) If $\varepsilon_{pc}\simeq\varepsilon_{ec}\simeq\varepsilon_{cc}$,
the event number of $\bar{\nu}_e$ observed through the channel
of the inverse beta decay is much larger than that of
the neutrino-electron reactions and neutrino-carbon reactions;

(iii) For the inverse beta decay, the event number of $\bar{\nu}_e$ calculated
with the neutrino energy spectra described by the Fermi-Dirac distribution
is larger than that by the ``beta fit" distribution;
however, for the neutrino-carbon reactions, the event numbers of different flavor neutrinos
calculated with the neutrino energy spectra described by the Fermi-Dirac distribution
are all smaller than that by the ``beta fit" distribution;

(iv) For the neutrino-electron reactions and neutrino-carbon reactions,
the event numbers of $\nu_e$ and $\bar{\nu}_e$ are larger than that of
$\nu_x$ and $\bar{\nu}_x$;

(v) More precise values of
$T_{\alpha}$ and $\eta_{\alpha}$ ($\langle E_{\alpha}\rangle$ and $\beta_{\alpha}$) will help obtain more reliable
event number ranges of SN neutrinos \cite{Mirizzi1, Tamborra1}.

(vi) Until the completion of the design of neutrino detector at
CSNS, the detector efficiency will be given and more accurate event number ranges of SN neutrinos
can be gained.
\vspace{0.3cm}

In the next section, the numerical calculation method of SN neutrino detection
on the Earth will be applied for some other spallation neutron sources at GeV energy range in the world.

\section{SN neutrino detection at other spallation neutron sources}

\begin{center}
\tabcaption{ \label{tab3} Summary of the proposed neutrino detectors at some current spallation neutron
sources at GeV energy range. }
\footnotesize
\begin{tabular*}{80mm}{@{\extracolsep{\fill}}c|ccccc}
\toprule Detector   & Material   & Mass   &  Depth  & Target numbers\\
           &            & (kton) &  (km) &  \\
 \hline
                             &             &           &          &  $N_p$: $7.62\times10^{31}$  \\
 $\nu$SNS \cite{Elnimr1}    &   CH$_2$    &  0.886    &  0.006   &  $N_e$: $3.05\times10^{32}$   \\
                             &             &           &          &  $N_C$: $3.81\times10^{31}$   \\
 \hline
                         &                 &           &          &  $N_p$: $3.35\times10^{34}$   \\
 $\nu$ESS \cite{ESS2}   &   H$_2$O        &  500      &  1.0     &  $N_e$: $1.67\times10^{35}$   \\
                         &                 &           &          &  $N_O$: $1.67\times10^{34}$   \\
\bottomrule
\end{tabular*}%
\end{center}

In the history of neutrino research, there are many famous neutrino experiments
which were based on spallation neutron sources and gained very important achievements \cite{Burman1},
 such as the Liquid Scintillator Neutrino Detector (LSND)
\cite{LSND1} at the Los Alamos Meson Physics Facility (LAMPF), the Karlsruhe Rutherford
Medium Energy Neutrino experiment (KARMEN) \cite{KARMEN1, KARMEN4} at the Spallation Neutron Source
of Rutherford Appleton Laboratory (ISIS) \cite{ISIS1} and so on.
Since the 21st century, some new spallation neutron
sources have been finished or under construction, and they
may be used for accelerator neutrino experiments in the future, such as the Spallation
Neutron Source at Oak Ridge National Laboratory (SNS) \cite{SNS1},
CSNS \cite{CSNS2}, the European Spallation Neutron Source (ESS)
\cite{ESS1} and so on.
In Table 3, we list the material of liquid scintillator, the detector masses, the
underground depth of the detectors, and the target numbers of the proposed neutrino detectors at
some current spallation neutron sources at GeV energy range in the world. Similar to the proposed
neutrino detector at CSNS discussed in the above section, these neutrino detectors can also be used
for observing SN neutrinos, and some information about the neutrino mixing parameters and explosion
mechanism of SN may be gained.

For the neutrino detector at SNS, there are three reaction channels which can be
used to detect SN neutrinos: the inverse beta decay,
the neutrino-electron reactions, and the neutrino-carbon reactions, and their effective cross sections
are given in Table 1. For the neutrino detector at ESS,
there are also three reaction channels which can be
used to detect SN neutrinos: the inverse beta decay,
the neutrino-electron reactions, and the neutrino-oxygen reactions. The effective cross sections
for the inverse beta decay and neutrino-electron reactions are given in Table 1, and the total effective
cross sections for the neutrino-oxygen reactions are given as follows
\cite{OCS1, OCS2, OCS3}:
\begin{eqnarray}
  \langle\sigma(^{16}O(\nu_e,e^-)^{16}F^*)\rangle =  1.91\times10^{-43}cm^2, \nonumber \\
  \langle\sigma(^{16}O(\Bar{\nu}_e,e^+)^{16}N^*)\rangle =   1.05\times10^{-42}cm^2, \nonumber\\
  \langle\sigma(^{16}O(\nu_{x},\nu_{x}')^{16}O^*)\rangle = 5.90\times10^{-42}cm^2 , \label{OSC} \\
  \langle\sigma(^{16}O(\Bar{\nu}_{x},\Bar{\nu}_{x}')^{16}O^*)\rangle = 4.48\times10^{-42}cm^2. \nonumber
\end{eqnarray}

The effective cross sections of the charged-current interaction in Eq. (10) are given for
SN neutrinos without oscillations. When neutrino oscillations are taken into account, the oscillations of higher
energy $\nu_x$ into $\nu_e$ result in an increased event rate
since the expected $\nu_e$ energies are just at or below the
reaction threshold. This leads to an increase by a
factor of 71.7 for the efficiency cross section
$\langle\sigma(^{16}O(\nu_e,e^-)^{16}F)\rangle$.
Similarly, the efficiency cross section
$\langle\sigma(^{16}O(\Bar{\nu}_e,e^+)^{16}N)\rangle$ is growing
by a factor of 9.2.

Similar to the CSNS detector, in order to calculate the event numbers of SN neutrinos more accurately,
for the SNS detector, we choose to use the detector efficiency $\varepsilon_{ps}$ for the inverse beta decay, $\varepsilon_{es}$ for the neutrino-electron reactions, $\varepsilon_{cs}$ for the neutrino-carbon reactions,
respectively; for the ESS detector, we choose to use the detector efficiency $\varepsilon_{pe}$ for the inverse beta decay, $\varepsilon_{ee}$ for the neutrino-electron reactions, $\varepsilon_{oe}$ for the neutrino-oxygen reactions, respectively.

By using Eqs. (\ref{FD})-(\ref{Ntotal}), the event numbers of SN neutrinos detected
at SNS and ESS can be calculated. The numerical results calculated with the neutrino
energy spectra described by the Fermi-Dirac
distribution and ``beta fit" distribution are both shown in Table 4.
In the table, the total event number ranges of SN neutrinos observed through
various reaction channels: the inverse beta decay, the neutrino-electron reactions,
the neutrino-carbon reactions, and the neutrino-oxygen reactions,
are given. It can be found that:

(i) If the efficiencies of different detectors are similar,
then the total event number of SN neutrinos detected at ESS is much larger than that
at CSNS and SNS;

(ii) The total event numbers of SN neutrinos detected in the case of
inverted hierarchy are larger than that of normal hierarchy;

(iii) If the detector efficiencies of different reaction channels are similar,
then the total event number of SN neutrinos observed through the channel
of the neutrino-electron reactions is much smaller than that of
the inverse beta decay, the neutrino-carbon reactions, and the neutrino-oxygen reactions;

(iv) For the inverse beta decay, the total event number of SN neutrinos calculated
with the neutrino energy spectra described by the Fermi-Dirac distribution
is larger than that by the ``beta fit" distribution;
however, for the neutrino-carbon reactions and neutrino-oxygen reactions, the total event
numbers of SN neutrinos calculated with the neutrino energy spectra described
by the Fermi-Dirac distribution are both smaller
than that by the ``beta fit" distribution.

\vspace{0.3cm}

\end{multicols}

\begin{center}
\tabcaption{ \label{tab4}  Summary of the total event number ranges of
SN neutrinos detected in various reactions at SNS and ESS. ``Range (FD)" (``Range (BF)")
stands for the event number ranges of SN neutrinos calculated by using the Fermi-Dirac (``beta fit") distribution,
``[x, y]$\varepsilon$" stands for the event number
range satisfied $x\times\varepsilon\leqslant N \leqslant y\times\varepsilon$ where $\varepsilon$ is the
detector efficiency.}
\footnotesize
\begin{tabular*}{170mm}{@{\extracolsep{\fill}}c|ccccc}
\toprule Detector  &  Hierarchy  & Reaction    & Range (FD) & Range (BF)\\
  \hline
           &   Normal        &  $\bar{\nu}_ep$         &  [410.21, 566.47]$\varepsilon_{ps}$    &
           [244.33, 350.02]$\varepsilon_{ps}$  \\
           &                 &  $\nu e^-$              &  [14.88, 15.25]$\varepsilon_{es}$      &
           [15.15, 15.26]$\varepsilon_{es}$  \\
  $\nu$SNS &                 &  $\nu^{12}C$            &  [79.13, 118.26]$\varepsilon_{cs}$     &
           [142.95, 174.88]$\varepsilon_{cs}$  \\
           &  Inverted       &  $\bar{\nu}_ep$         &  [493.13, 707.65]$\varepsilon_{ps}$    &
           [281.85, 391.62]$\varepsilon_{ps}$     \\
           &                 &  $\nu e^-$              &  [15.39, 15.71]$\varepsilon_{es}$      &
           [15.44, 15.58]$\varepsilon_{es}$ \\
           &                 &  $\nu^{12}C$            &  [97.02, 141.01]$\varepsilon_{cs}$     &
           [170.92, 205.36]$\varepsilon_{cs}$  \\
  \hline
           &   Normal        &  $\bar{\nu}_ep$         &     [$1.80\times10^5$, $2.49\times10^5$]$\varepsilon_{pe}$  & [$1.07\times10^5$, $1.54\times10^5$]$\varepsilon_{pe}$ \\
           &                 &  $\nu e^-$              &     [$8.15\times10^3$, $8.35\times10^3$]$\varepsilon_{ee}$  &  [$8.30\times10^3$, $8.35\times10^3$]$\varepsilon_{ee}$  \\
  $\nu$ESS &                 &  $\nu^{16}O$            &     [$4.73\times10^4$, $7.18\times10^4$]$\varepsilon_{oe}$  &
           [$8.72\times10^4$, $1.07\times10^5$]$\varepsilon_{oe}$   \\
           &  Inverted       &  $\bar{\nu}_ep$         &     [$2.17\times10^5$, $3.11\times10^5$]$\varepsilon_{pe}$  & [$1.24\times10^5$, $1.72\times10^5$]$\varepsilon_{pe}$\\
           &                 &  $\nu e^-$              &     [$8.43\times10^3$, $8.61\times10^3$]$\varepsilon_{ee}$  & [$8.45\times10^3$, $8.53\times10^3$]$\varepsilon_{ee}$  \\
           &                 &  $\nu^{16}O$            &     [$5.89\times10^4$, $8.56\times10^4$]$\varepsilon_{oe}$  &   [$1.04\times10^5$, $1.25\times10^5$]$\varepsilon_{oe}$ \\
\bottomrule
\end{tabular*}%
\end{center}

\begin{multicols}{2}

\section{Summary and discussion}

In this paper, the SN neutrino detection on the Earth was studied.
While considering all effects: the SN shock effects, the MSW effects,
the neutrino collective effects, and the Earth matter
effects, the detection of SN neutrinos on the Earth was studied.
Then, the event number ranges of different flavor SN neutrinos
observed through various reaction channels at CSNS: the inverse beta decay,
the neutrino-electron reactions and the neutrino-carbon reactions, were calculated
with the neutrino energy spectra described by the Fermi-Dirac distribution and
``beta fit" distribution respectively.

Applying the numerical calculation method of SN neutrino detection on the Earth
to some other spallation neutron sources (SNS and ESS) at GeV energy range,
and the total event number ranges of SN neutrinos detected in different reactions
channels (SNS: the inverse beta decay, the neutrino-electron reactions,
the neutrino-carbon reactions; ESS: the inverse beta decay,
the neutrino-electron reactions, the neutrino-oxygen reactions) were calculated.

In the future, after the completion of the design of neutrino detector at
spallation neutron sources (CSNS, SNS, ESS), the detector efficiency will be given and
more accurate event number ranges of SN neutrinos can be gained.
Furthermore, it is known that more precise values of
$T_{\alpha}$ and $\eta_{\alpha}$ ($\langle E_{\alpha}\rangle$ and $\beta_{\alpha}$)
will give more reliable event number ranges of SN neutrinos.

\vspace{10mm}

\acknowledgments{The authors would like to thank S. Wang and S.-J. Ding for
helpful discussions and support.}

\end{multicols}

\vspace{10mm}

\vspace{-1mm}
\centerline{\rule{80mm}{0.1pt}}
\vspace{2mm}

\begin{multicols}{2}

\end{multicols}

\clearpage

%\end{CJK*}
\end{document}